\documentclass[a4paper]{article}

\usepackage{INTERSPEECH2022}

\usepackage{graphicx}
\usepackage{amsmath}
\usepackage{amssymb}
\usepackage{booktabs}
\usepackage{multirow}
\usepackage{enumitem}
\usepackage[hang,flushmargin]{footmisc}
\usepackage[hyphens]{url}

\usepackage{hyperref}
\hypersetup{
colorlinks=true,
linkcolor=blue,
citecolor=blue,
urlcolor=blue,
}

\title{Chain-based Discriminative Autoencoders for Speech Recognition}

\name{Hung-Shin Lee, Pin-Tuan Huang, Yao-Fei Cheng, and Hsin-Min Wang}
\address{
Institute of Information Science, Academia Sinica}
\email{hungshinlee@gmail.com}

\begin{document}

\maketitle

\begin{abstract}
In our previous work, we proposed a discriminative autoencoder (DcAE) for speech recognition. DcAE combines two training schemes into one. First, since DcAE aims to learn encoder-decoder mappings, the squared error between the reconstructed speech and the input speech is minimized. Second, in the code layer, frame-based phonetic embeddings are obtained by minimizing the categorical cross-entropy between ground truth labels and predicted triphone-state scores. DcAE is developed based on the Kaldi toolkit by treating various TDNN models as encoders. In this paper, we further propose three new versions of DcAE. First, a new objective function that considers both categorical cross-entropy and mutual information between ground truth and predicted triphone-state sequences is used. The resulting DcAE is called a chain-based DcAE (c-DcAE). For application to robust speech recognition, we further extend c-DcAE to hierarchical and parallel structures, resulting in hc-DcAE and pc-DcAE. In these two models, both the error between the reconstructed noisy speech and the input noisy speech and the error between the enhanced speech and the reference clean speech are taken into the objective function. Experimental results on the WSJ and Aurora-4 corpora show that our DcAE models outperform baseline systems.
\end{abstract}

\section{Introduction}
With the advent and breakthroughs of deep learning, current automatic speech recognition (ASR) systems perform remarkably well in relatively clean conditions. However, due to the mismatch between training and test conditions, noise can greatly degrade performance in practical real-world application scenarios. Numerous techniques have been proposed to reduce the adverse effects of noise on acoustic modeling caused by interfering background events (such as babble or street noise), reverberation generated from non-ideal room acoustics, or channel distortion caused by different microphone characteristics \cite{Mitra2017}. These techniques mainly fall into two categories: multi-condition training (MCT) and speech enhancement (SE)-based pre-processing \cite{Yu2013,Estimator1985,Nakatani2010,Xu2014,Maas2012}.

MCT is widely adopted to achieve model robustness against additive noise, channel mismatch \cite{Seltzer2013,Du2014,Yu2013}, and reverberation \cite{Couvreur2000,Hsiao2015}. As stated in \cite{Yin2015}, adding extra noisy speech to the training data is a form of regularization that can provide better generalization. In addition, MCT has been shown to provide better performance than SE-based front-ends under unseen degradation conditions \cite{Seltzer2013,Du2014}. However, the performance under unseen conditions still lags behind the performance obtained in seen conditions \cite{Du2014,Qian2015}.

On the other hand, empirical experience also shows that SE as a front-end can improve ASR performance. Among the achievements in integrating SE into ASR systems \cite{Xu2014}, Narayanan and Wang inherited the SE model in \cite{Wang2013} and firstly proposed joint training of a time-frequency (T-F) mask-based front-end and a deep neural network (DNN)-based acoustic model back-end \cite{Narayanan2014}. The front-end was used to estimate clean features from noisy features, and the back-end was used to classify phoneme labels. They were connected as a combined neural network, and the model parameters were simultaneously optimized with back-propagation. Related extensions can be seen in \cite{Narayanan2015,Narayanan2015a,Gao2015}. In addition, Soni and Panda recently proposed another version of this type of framework in \cite{Soni2019}, which can be used without parallel clean-noisy speech training data and without pre-trained feature enhancement modules.

The above topology of integrated SE and phoneme classification can be simply expressed by \textit{input features} / \textit{enhancement block} / \textit{classification block} / \textit{output targets}. A detailed description can be seen in Fig. 2 of \cite{Fujimoto2019}, which shows how these blocks are trained by joint training (JT) and multi-task learning (MTK). In this study, we extend our previous work, i.e., discriminative autoencoder (DcAE) \cite{Huang2019,Yang2017}, and propose new neural architectures, in which the \textit{classification block} is wrapped in the \textit{enhancement block} for robust speech recognition. A denoising autoencoder has been a popular DNN structure that can learn the mapping between noisy and clean speech in the frequency or time domains \cite{Pascual2017,Rethage2018,Tawara2018,Keisuke2020} from single- or multi-channel observations \cite{Lu2013,Araki2015}. 

\begin{figure*}
\centering
\includegraphics[width=1.0\textwidth]{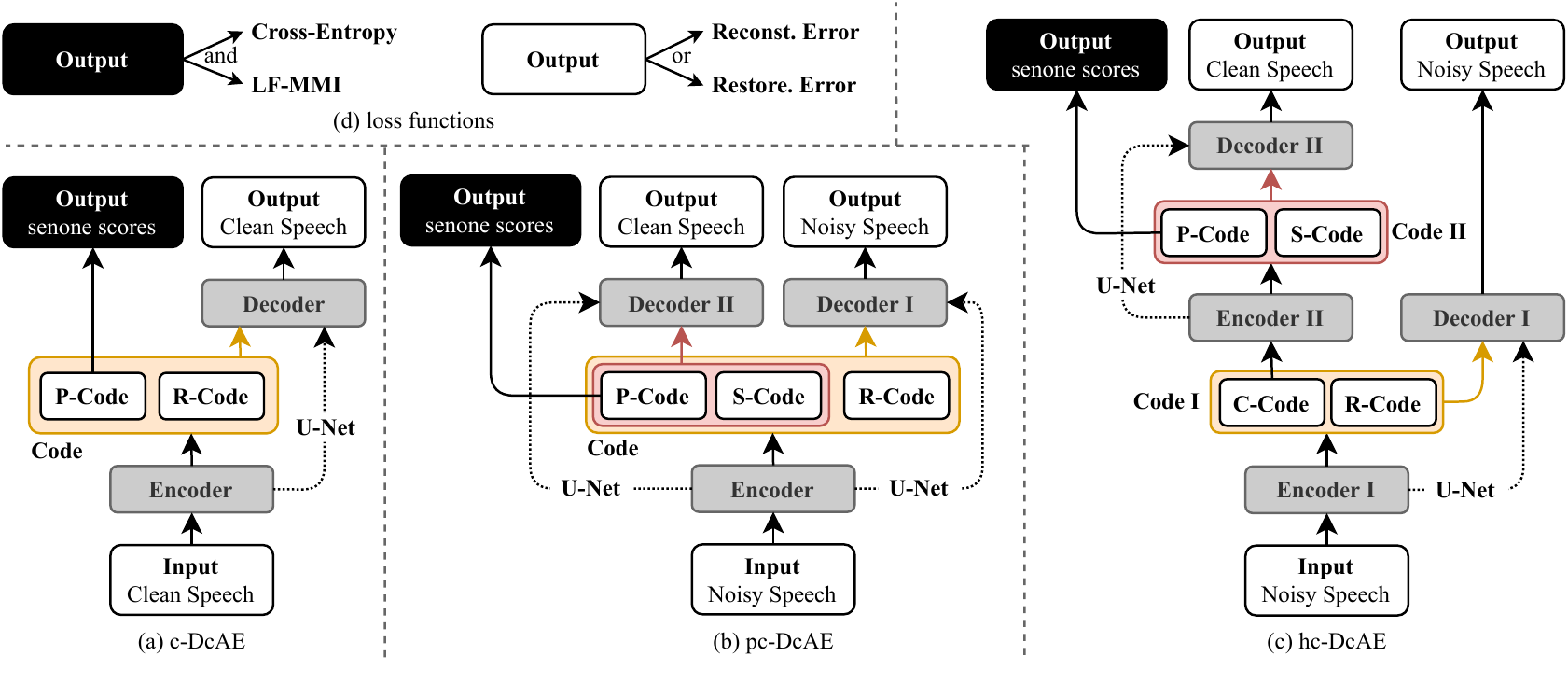}
\vspace{-25pt}
\caption{Different versions of DcAE. (a) Chain-based DcAE (c-DcAE), where the code layer is divided into P-Code (phoneme-aware code) and R-Code (residual code). P-Code directly contributes to senone scores. (b) Parallel c-DcAE (pc-DcAE), where the predicted clean speech presumably comes from P-Code and S-Code (speaker-aware code) through Decoder II, and the predicted noisy speech presumably comes from P-Code, S-Code, and R-Code through Decoder I. (c) Hierarchical c-DcAE (hc-DcAE), where C-Code (clean speech code) and R-Code are used to generate noisy speech by Decoder I, and C-Code is further factorized by Encoder II into P-Code and S-Code, which are used to generate clean speech by Decode II. (d) Two types of Output and the associated losses.}
\label{fig:structure}
\vspace{-10pt}
\end{figure*}

In summary, our work has at least three major contributions:
\begin{enumerate}[label=\arabic*),wide,labelindent=0pt]
\item Compared to frame-level cross-entropy, few implementations have introduced the lattice-free maximum mutual information (LF-MMI) criterion into common autoencoder-based ASR or into \textit{joint training} for robust acoustic modeling.
\item Two variants of DcAE are proposed to integrate SE into acoustic modeling. What they have in common is that the enhancement block not only removes noise, but also restores the original signal. Their major difference lies in the way that the enhancement block reconstructs noisy and clean speech: in parallel or hierarchically. The new DcAE can not only extract cleaner and more discriminative embeddings for triphone-state (senone) classification, but also improve generalization. The decoder layers in DcAE can be regarded as a kind of regularizer, as described in \cite{Huang2019}.
\item Compared with the baseline model without decoder and code layers, DcAE achieves relative word error rate (WER) reductions of 11.09\% and 10.28\% in {\ttfamily dev93} and {\ttfamily eval92} on the WSJ corpus. Experimental results on the Aurora-4 corpus show that DcAE achieves a relative WER reduction of 5.86\% over the baseline model trained with MCT.
\end{enumerate}

\section{Proposed Neural Structures}

An autoencoder is a learning model designed to convert an input into a reconstructed output with minimal distortion. The embedded internal representation, i.e., the code layer, contains the most critical information that can be used to minimize the reconstruction error. In our previous work, in order to obtain more accurate acoustic scores for ASR, we divided the code layer into two sub-representations, the phoneme-aware code (P-Code) and the residual code (R-Code), to form a basic version of DcAE \cite{Huang2019}. P-Code contains phonetic information and is connected to the original output layer of the acoustic model of the ASR system. R-Code carries information unrelated to phonetic content, which may include environmental noise, speaker identity, and any other information deemed useless for the ASR task. P-Code and R-Code are concatenated as input to the decoder to reconstruct the original acoustic frame. This version of DcAE was implemented with the time-delay neural network (TDNN) \cite{Waibel1989,Peddinti2015} and its combination with long short-term memory (TDNN-LSTM) \cite{Peddinti2017} for acoustic modeling. The cross-entropy (CE) was used as the ASR loss. In this section, we introduce the chain-based DcAE and its two variants, which can make acoustic models more robust in noisy environments when clean-noisy utterance pairs are sufficient for model training.

\subsection{Chain-based DcAE (c-DcAE)}
There are two main requirements for a model to be called ``chain-based'' in the sense of the Kaldi toolkit\footnote{\url{https://kaldi-asr.org/doc/chain.html}} \cite{Vesely2013,Povey2016}. First, due to the reduced frame rate of the neural network output, an unconventional hidden Markov model (HMM) topology that allows traversal of the HMM in one state is used. Second, the sequence-level objective function, i.e., the log-probability of the correct sequence realized by MMI, is jointly optimized with CE for model training. As shown in Fig. \ref{fig:structure}(a), the senone scores (the black box) are compared with the reference senone labels using CE and LF-MMI in the objective function.

\subsection{Parallel Chain-based DcAE (pc-DcAE)}
The first version of robust DcAE is the parallel chain-based DcAE (pc-DcAE). As shown in Fig. \ref{fig:structure}(b), the architecture of pc-DcAE can be simply described as \textit{Input / Encoder / Code / Decoder $\times$2 / Output $\times$3}. A noisy acoustic frame and its corresponding speaker embedding (e.g., i-vector) are fed into the encoder, which can be implemented with various neural networks. In the code layer (Code), the nodes are divided into three parts. The phoneme-aware code (P-Code) that is designed to contain as much phonetic information as possible is directly connected to the senone scores. S-Code contains the speaker information while R-Code carries other remaining information, especially unwanted noise. Decoder II is used to reconstruct the corresponding clean acoustic frame of the input noisy acoustic frame according to the concatenated P-Code and S-Code. Decoder I is used to reconstruct the noisy acoustic frame according to the concatenated P-Code, S-Code, and R-Code. Both decoders are implemented with fully-connected networks.

Note that S-Code is not formulated by training with actual speaker labels, but it would make more sense if associated with a speaker-related loss. S-Code represents speaker information, since ideal clean speech can be viewed as a composite signal present only in orthogonal phonetic and speaker dimensions. 

\begin{table*}[ht]
\centering
\caption{WERs (relative changes over Baseline) (\%) on WSJ. Both Baseline and Encoder of c-DcAE use TDNN-F. Spec denotes ratio of code sizes/depth of Decoder/weight for U-Net/no. of training epochs for warm start in c-DcAE (cf. Fig. \ref{fig:structure}). N means ``not applied''.}
\vspace{-5pt}
\label{tab:wsj}
\begin{tabular*}{\linewidth}{@{\extracolsep{\fill}} l@{\hspace{6pt}}|c@{\hspace{5pt}}cc@{\hspace{5pt}}c|c@{\hspace{5pt}}cc@{\hspace{5pt}}c}
\toprule
\multirow{2}{*}{\textbf{Model}} & \multicolumn{4}{c|}{\textbf{Flat Start}} & \multicolumn{4}{c}{\textbf{Warm Start}} \\
\cmidrule{2-5}
\cmidrule{6-9}
& {\ttfamily dev93} & \textbf{Spec} & {\ttfamily eval92} & \textbf{Spec} & {\ttfamily dev93} & \textbf{Spec} & {\ttfamily eval92} & \textbf{Spec} \\
\cmidrule{1-1}
\cmidrule{2-3}
\cmidrule{4-5}
\cmidrule{6-7}
\cmidrule{8-9}
Baseline & 4.42 & - & 2.53 & - & 4.42 & - & 2.53 & - \\
\midrule
c-DcAE & 3.96 (10.41) & 0.5/2/N/N & 2.30 (9.09) & 0.5/6/N/N & 4.02 (9.05) & 0.5/2/N/4 & 2.34 (7.51) & 0.5/2/N/4 \\
+ U-Net (concat) & 3.96 (10.41) & 0.5/5/N/N & 2.30 (9.09) & 0.5/5/N/N & 4.02 (9.05) & 0.5/5/N/6 & 2.41 (4.74) & 0.5/5/N/4 \\
+ U-Net (sum) & 4.09 (7.47) & 0.5/2/0.8/N & 2.29 (9.49) & 0/3/0.3/N & 4.04 (8.60) & 0/3/0.3/6 & \bf 2.30 (9.09) & 0/3/0.3/4 \\
+ U-Net (diff+concat) & \bf 3.93 (11.09) & 0/3/0.3/N & \bf 2.27 (10.28) & 0.5/3/0.5/N &\bf 4.01 (9.28) & 0.5/3/0.5/5 & 2.36 (6.72) & 0/3/0.3/4 \\
\bottomrule
\end{tabular*}
\vspace{-10pt}
\end{table*}

\subsection{Hierarchical Chain-based DcAE (hc-DcAE)}
The second version of robust DcAE is the hierarchical chain-based DcAE (hc-DcAE). As shown in Fig. \ref{fig:structure}(c), the architecture of hc-DcAE can be simply described as \textit{Input / Encoder / Code / Encoder / Code / Decoder / Output}, where the clean acoustic frame is hierarchically distilled from the noisy acoustic frame through two code layers. There are two autoencoders lurking in hc-DcAE. The first is \textit{Input (Noisy Speech) / Encoder I / Code I / Decoder I / Output (Noisy Speech)}. By restoring the original noisy speech, it ensures that the code layer (Code I) contains the salient components of the speech. The second is \textit{C-Code / Encoder II / Code II / Decoder II / Output (Clean Speech)}, where C-Code, the junction of the two autoencoders, is defined as the embedding of a clean acoustic frame. Similar to pc-DcAE, both decoders in hc-DcAE are implemented with fully-connected networks.

The main difference between pc-DcAE and hc-DcAE is twofold. In pc-DcAE, all codes are generated separately in one operation by the same encoder, while in hc-DcAE, each of the two encoders has its specific function. Therefore, the model capacity of hc-DcAE may be larger that of pc-DcAE. Second, the processing of clean speech in hc-DcAE during model training is deeper and more complex.

\subsection{Objective Functions}

In addition to common ASR losses including CE and LF-MMI, the objective function used in the proposed framework also involves reconstruction and restoration errors.

\subsubsection{Reconstruction and restoration errors}

Suppose our proposed model $\mathcal{M}$ contains deterministic mappings $f(\cdot)$ and $g(\cdot)$, which are responsible for observation reconstruction and restoration, respectively. Given a set of noisy data $\mathcal{X}$ and its corresponding clean data $\mathcal{Y}$ ready to go through the inference-generation processes $\mathcal{X} \overset{f}\rightarrow\mathcal{X}$ and $\mathcal{X} \overset{g}\rightarrow\mathcal{Y}$, the average reconstruction and restoration errors denoted by $\mathcal{L}_{rc}$ and $\mathcal{L}_{rs}$, respectively, are given by
\begin{equation}
\label{eq:rc}
\mathcal{L}_{rc}(\mathcal{X}) = \frac{1}{|\mathcal{X}|} \sum_{\mathbf{x}\in\mathcal{X}} \|f(\mathbf{x}) - \mathbf{x}\|^2_2,
\end{equation}
\begin{equation}
\label{eq:rs}
\mathcal{L}_{rs}(\mathcal{X}) = \frac{1}{|\mathcal{X}|} \sum_{\mathbf{x}\in\mathcal{X},\mathbf{y}\in\mathcal{Y}} \|g(\mathbf{x}) - \mathbf{y}\|^2_2,
\end{equation}
where $\|\cdot\|^2_2$ is the 2-norm operator, and $|\mathcal{X}|$ is the sample size. $f(\mathbf{x})$ and $g(\mathbf{x})$ correspond to Output (Noisy Speech) and Output (Clean Speech) shown in Fig. \ref{fig:structure}, respectively.

\subsubsection{ASR losses}
\label{sec:phonme-aware}
There are two ASR losses related to Output (senone scores) in Fig. \ref{fig:structure} in our acoustic models. One, denoted by $\mathcal{L}_{CE}$, is the CE between the distribution represented by the reference label and the predicted distribution. The other, denoted by $-\mathcal{F}_{LF-MMI}$, is the negative LF-MMI criterion between the distributions of predicted and reference word sequences. See \cite{Vesely2013,Povey2016,Hadian2018} for details on LF-MMI training (also known as {\ttfamily chain} modeling).

By combining the criterion based on noisy feature reconstruction, clean feature restoration, phoneme-aware CE, and sequence-level LF-MMI, the objective function becomes
\begin{equation}
\label{eq:final_obj}
\mathcal{L}=-\mathcal{F}_{LF-MMI}+5\times\mathcal{L}_{CE}+\alpha\mathcal{L}_{rc}+\beta\mathcal{L}_{rs},
\end{equation}
where the weight 5 of $\mathcal{F}_{CE}$ follows most recipes for {\ttfamily chain} modeling. $\alpha$ and $\beta$ are weights to increase or decrease the regularization strength of $\mathcal{L}_{rc}$ and $\mathcal{L}_{rs}$, respectively. They were determined heuristically in our experiments.

\section{Experiments and Results}
\subsection{Datasets: WSJ and Aurora-4}
We evaluated c-DcAE on clean speech recognition using the Wall Street Journal (WSJ) dataset \cite{Paul1992}, which consists of WSJ0 (LDC93S6B) and WSJ1 (LDC94S13B). The training set (called {\ttfamily si284} in most Kaldi recipes) consists of 81 hours speech, and we used {\ttfamily dev93} and {\ttfamily eval92} as test sets.

We evaluated pc-DcAE and hc-DcAE on robust speech recognition using the Aurora-4 dataset. Aurora-4 \cite{Parihar2002} is a medium-level vocabulary task. The transcriptions were based on WSJ0 (LDC93S6B) \cite{Paul1992}. The dataset contains 16 kHz speech data with additive noise and linear convolutional channel distortion, which were synthetically introduced into clean speech. The training set contains 7,138 utterance pairs from 83 speakers, where each pair consists of clean speech and speech corrupted by one of six different noises (i.e., street, train station, car, babble, restaurant, and airport) at 10--20 dB SNR. The test set was generated by the same types of noise and microphones and was grouped into four subsets: clean, noisy, clean with channel distortion, and noisy with channel distortion, which are referred to as A, B, C, and D, respectively. The speech data were recorded by two microphones, one of which was a near-field microphone. The speech recorded by this microphone can be considered clean and noise-free. In contrast, the speech recorded by the second microphone can be considered noisy speech.

\begin{table*}[ht]
\centering
\caption{WERs (relative changes over Baseline) (\%) on Aurora-4. Spec denotes depth of Decoder I/depth of Decoder II/weight for U-Net/depth of Encoder II (cf. Fig. \ref{fig:structure}). N means ``not applied''. The total depth of Encoders I and II is equal to the depth of Baseline.}
\vspace{-5pt}
\label{tab:aurora4}
\begin{tabular}{lccccccccccccc}
\toprule
{\bf Encoder} & {\bf Method} & {\bf U-Net} & {\bf A} & {\bf B} & {\bf C} & {\bf D} & {\bf Average} & {\bf Spec} \\
\midrule
\midrule
\multirow{9}{*}{TDNN-F} & Baseline & - & 1.91 & 3.69 & 3.42 & 10.31 & 6.38 & - \\
\cmidrule{2-9}
& \multirow{4}{*}{pc-DcAE} & - & 1.70 (10.99) & 3.50 (5.15) & 3.16 (7.60) & 10.22 (0.87) & 6.23 (2.35) & 5/5/N/N \\
&& sum & 1.66 (13.09) & \bf 3.47 (5.96) & 3.21 (6.14) & 10.29 (0.19) & 6.25 (2.04) & 3/3/1/N \\
&& concat & 1.81 (5.24) & 3.63 (1.63) & 3.31 (6.14) & 10.37 (-0.58) & 6.37 (0.16) & 3/3/N/N \\
&& diff+concat & 1.70 (10.99) & 3.53 (4.34) & \bf 3.03 (11.40) & 10.25 (0.58) & 6.24 (2.19) & 5/5/1/N \\
\cmidrule{2-9}
& \multirow{4}{*}{hc-DcAE} & - & \bf 1.61 (15.71) & 3.50 (5.15) & 3.14 (8.19) & \bf 10.19 (1.16) & 6.21 (2.66) & 3/3/N/1 \\
&& sum & 1.66 (13.09) & 3.53 (4.34) & 3.29 (3.80) & 10.36 (-0.48) & 6.31 (1.10) & 4/4/1/1 \\
&& concat & 1.77 (7.33) & 3.55 (3.79) & 3.42 (0.00) & 10.29 (0.19) & 6.30 (1.25) & 3/3/N/1 \\
&& diff+concat & 1.74 (8.90) & 3.55 (3.79) & 3.06 (10.53) & 10.04 (2.62) & \bf 6.17 (3.29) & 3/3/0.8/2 \\
\midrule
\multirow{9}{*}{\shortstack[1]{CNN-\\TDNNF}} & Baseline & - & 1.72 & 3.19 & 3.40 & 9.48 & 5.80 & - \\
\cmidrule{2-9}
& \multirow{4}{*}{pc-DcAE} & - & \bf 1.57 (8.72) & \bf 3.19 (0.00) & 3.03 (10.88) & 9.04 (4.64) & 5.57 (3.97) & 5/5/N/N \\
&& sum & 1.63 (5.23) & \bf 3.19 (0.00) & 3.10 (8.82) & 8.93 (5.80) & 5.53 (4.66) & 5/5/1/N \\
&& concat & 1.64 (4.65) & 3.20 (-0.31) & 2.86 (15.88) & 9.05 (4.54) & 5.57 (3.97) & 5/5/N/N \\
&& diff+concat & 1.59 (7.56) & 3.26 (-2.19) & 2.99 (12.06) & 8.83 (6.86) & 5.51 (5.00) & 4/4/0.5/N \\
\cmidrule{2-9}
& \multirow{4}{*}{hc-DcAE} & - & 1.63 (5.23) & 3.21 (-0.63) & \bf 2.67 (21.47) & \bf 8.82 (6.96) & \bf 5.46 (5.86) & 4/4/N/3 \\
&& sum & \bf 1.57 (8.72) & 3.27 (-2.51) & 2.86 (15.88) & 8.97 (5.38) & 5.56 (4.14) & 5/5/0.8/3 \\
&& concat & 1.72 (0.00) & \bf 3.19 (0.00) & 3.10 (8.82) & 8.94 (5.70) & 5.54 (4.48) & 4/4/N/3 \\
&& diff+concat & 1.59 (7.56) & 3.26 (-2.19) & 2.93 (13.82) & 8.91 (6.01) & 5.54 (4.48) & 5/5/1/1 \\
\bottomrule
\end{tabular}
\vspace{-10pt}
\end{table*}

\subsection{Model Settings}
All ASR systems were developed using a Gaussian mixture model (GMM)/DNN/HMM-based process based on a standard Kaldi recipe\footnote{\url{https://github.com/kaldi-asr/kaldi/blob/master/egs/{wsj/s5/run.sh,aurora4/s5/run.sh}}}. The standard four-gram and trigram language models were used in the WSJ and Aurora-4 tasks, respectively.

Our DcAE models were built on top of two acoustic models, factorized TDNN (TDNN-F) \cite{Povey2018} and its convolutional neural network-based variant (CNN-TDNNF). TDNN-F consists of 13 hidden layers, each containing 1024-dimensional hidden nodes and 128-dimensional linear bottleneck nodes. In addition, the parameters of the final layer are also factorized using a 192-node linear bottleneck layer. CNN-TDNNF consists of 15 layers. The first six layers are convolutional and the remaining nine layers are TDNN-F. The time offsets and height offsets were set to ``-1, 0, 1'' for all convolutional layers. The numbers of filters for the six convolutional layers were 48, 48, 64, 64, 64, and 128, respectively. The parameters of the TDNN-F layers were set to the same value as TDNN-F.

Forty high-resolution MFCCs (obtained by discrete cosine transform of 40 Mel-frequency bins and normalized by utterance-based mean subtraction) and the 100-dimensional i-vector \cite{Dehak2011} were used as frame input for all systems.

\subsection{Results}
The structure of c-DcAE provides an opportunity for model pre-training using acoustic features only (cf. warm start in Table \ref{tab:wsj}). Experimental results on WSJ show that training from scratch using acoustic features and phoneme labels (cf. flat start) works better. However, we found that due to some training schemes used in Kaldi, the training error defined in Eq. \ref{eq:rc}, although eventually converged, did not drop much. Although worse than training from scratch, warm start still shows its potential and superiority to the baseline by reducing the WER by 9.28\% and 9.09\% in {\ttfamily dev93} and {\ttfamily eval92}, respectively. It is worth noting that it should make more sense to pre-train on large amounts of unlabeled data. We will investigate this further in the future. Due to space limitations, we do not present the results of CNN-TDNNF in Table \ref{tab:wsj}, but instead compare various U-Net operations \cite{Ronneberger2015} applied in c-DcAE (cf. Fig. \ref{fig:structure}). As shown in Table \ref{tab:wsj}, c-DcAE+U-Net(diff+concat), where the layers in Decoder were concatenated by the difference between the layers of Encoder and Decoder, achieved the best performance and reduced the WER by 11.09\% and 10.28\% in {\ttfamily dev93} and {\ttfamily eval92}, respectively. Note that the weight for U-Net denotes the scaling strength coming from Encoder (for ``sum'') or the difference between Encoder and Decoder (for ``diff+concat''), and the ratio of code sizes is defined by ``$\text{the size of R-Code} / \text{the size of P-Code}$''. The size of P-Code was fixed to 1,024.

In robust speech recognition experiments on Aurora-4, the sizes of P-Code, S-Code and R-Code were set to 1,024 in pc-DcAE and hc-DcAE. From Table \ref{tab:aurora4}, we can see that pc-DcAE (diff+concat) achieved the best performance and reduced the WER by 11.40\% in Set C (channel-distorted clean speech), when TDNN-F was used as Encoder. The CNN-TDNNF-based hc-DcAE could even reduce the WER by 21.47\% in Set C. This result shows that DcAE is more robust to channel mismatch than Baseline. An interesting observation is that both pc-DcAE and hc-DcAE had less improvement over Baseline under noisy conditions (Sets B and D). CNN-TDNNF-based pc-DcAE and hc-DcAE performed even worse than Baseline in Set B. A possible reason might be that CNN-TDNNF itself is robust enough to additive noise, since the first few convolutional layers can filter out non-speech sounds to some extent. This somewhat conforms with the fact that most Transformer-based end-to-end ASR systems tend to use CNN for front-end subsampling \cite{Watanabe2018,Gulati2020}. Furthermore, the benefit that CNN-TDNNF brings to pc-DcAE and hc-DcAE is that all modifications by U-Net seem unnecessary. As can be seen from the last 4 rows of Table \ref{tab:aurora4}, naive hc-DcAE achieved the best performance with a relative reduction of 5.86\% in average WER. This result is also better than most studies in the literature thus far, except for CNN-Raw \cite{Loweimi2020}, which achieves a WER of 5.10\%\footnote{In CNN-Raw, the forced alignment of \textit{clean} speech was directly applied to the corresponding \textit{noisy} speech. Considering that \textit{clean} speech corresponding to \textit{noisy} speech is not always available, we performed forced alignment on \textit{clean} speech and \textit{noisy} speech separately.}.

\section{Conclusion}

In this paper, we have proposed a neural architecture, the chain-based discriminative autoencoder (c-DcAE), which successfully combines {\ttfamily chain} modeling and DcAE in acoustic modeling. For robust ASR, two variants of c-DcAE, namely pc-DcAE and hc-DcAE, are further proposed to process noisy and clean speech in parallel and hierarchical manners, respectively. Joint training involves ASR-related losses, especially lattice-free MMI, and feature reconstruction/restoration errors. Several types of connections using U-Net (like skip-connections) are introduced into the structure of DcAE. Another highlight of our DcAE-based models is that they do not require additional computation time than the baseline in the inference stage, as all decoders in DcAE can be removed at inference time. Our experimental results have shown that the proposed models outperform the baseline models on the WSJ and Aurora-4 tasks.

\vfill\pagebreak

\bibliographystyle{IEEEtran}
\bibliography{references.bib}
\end{document}